\documentclass[aps,preprintnumbers,nofootinbibt,referee, keywords]{revtex4}
\usepackage{graphicx}
\usepackage{amsmath}
\usepackage{bm}
\usepackage{color}

\def\be{\begin{equation}}
\def\ee{\end{equation}}
\def\bea{\begin{eqnarray}}
\def\eea{\end{eqnarray}}

\begin{document}

\title{Integrability cases for the anharmonic oscillator equation}
\author{Tiberiu Harko}
\email{t.harko@ucl.ac.uk}
\affiliation{Department of Mathematics, University College London, Gower Street, London
WC1E 6BT, United Kingdom}
\author{Francisco S. N. Lobo}
\email{flobo@cii.fc.ul.pt}
\affiliation{Centro de Astronomia e Astrof\'{\i}sica da Universidade de Lisboa, Campo
Grande, Ed. C8 1749-016 Lisboa, Portugal}
\author{M. K. Mak}
\email{mkmak@vtc.edu.hk}
\affiliation{Department of Physics and Center for Theoretical and Computational Physics,
The University of Hong Kong, Pok Fu Lam Road, Hong Kong, P. R. China}

\begin{abstract}
Using N. Euler's theorem on the integrability of the general anharmonic
oscillator equation \cite{12}, we present three distinct classes of general
solutions of the highly nonlinear second order ordinary differential
equation $\frac{d^{2}x}{dt^{2}}+f_{1}\left( t\right) \frac{dx}{dt}%
+f_{2}\left( t\right) x+f_{3}\left( t\right) x^{n}=0$. The first exact
solution is obtained from a particular solution of the point transformed
equation $d^{2}X/dT^{2}+X^{n}\left( T\right) =0$, $n\notin \left\{
-3,-1,0,1\right\} $, which is equivalent to the anharmonic oscillator
equation if the coefficients $f_{i}(t)$, $i=1,2,3$ satisfy an integrability
condition. The integrability condition can be formulated as a Riccati
equation for $f_{1}(t)$ and $\frac{1}{f_{3}(t)}\frac{df_{3}}{dt}$
respectively. By reducing the integrability condition to a Bernoulli type
equation, two exact classes of solutions of the anharmonic oscillator
equation are obtained.
\end{abstract}

\maketitle


\section{ Introduction}

The anharmonic oscillator is a physical system generalizing the simple
linear harmonic oscillator $\frac{d^{2}x}{dt^{2}}+\omega _{0}^{2}x(t)=0$,
where $x(t)$ is the position coordinate, $t$ is the time, and $\omega _{0}$
is the oscillation frequency. In general, the time evolution of the space
variable $x$ of the anharmonic oscillator is governed by the following
nonlinear second order differential equation \cite{7,12}
\begin{equation}
\frac{d^{2}x}{dt^{2}}+f_{1}\left( t\right) \frac{dx}{dt}+f_{2}\left(
t\right) x+f_{3}\left( t\right) x^{n}=f_{4}\left( t\right) ,  \label{1}
\end{equation}%
where $f_{i}\left( t\right) $, $i=1,2,3,4$, and $x$ are arbitrary real
functions of $t$ defined on a real interval $I\subseteq \Re $, with $%
f_{i}\left( t\right) $ and $x\left( t\right) $ $\in C^{\infty }(I)$. The
factors $f_{i}\left( t\right) $ are physically interpreted as follows: $%
f_{1}\left( t\right) $ is a damping factor; $f_{2}\left( t\right) $ is a
time dependent oscillation frequency coefficient; $f_{3}\left( t\right) $ is
the simplest possible anharmonic term; $f_{4}\left( t\right) $ is a forcing
term, and $n$ is a real constant \cite{7}. The equation of motion of the anharmonic
oscillator is strongly nonlinear, and when the anharmonicity term $%
f_{3}\left( t\right) x^{n}$ is small, its solutions can be obtained by using
perturbation theory. If the anharmonicity is large, then other numerical
techniques need to be implemented.

The anharmonic oscillator equation Eq.~(\ref{1}) with specific values of the
exponent $n$ can be used to model many different physical systems. For $%
n=3/2 $, one obtains the Thomas and Fermi atomic model \cite{2,3}, while the
case $n=-3$ corresponds to the Ermakov \cite{4}, or Pinney \cite{5},
equation. For $n=-1$ one  obtains the Brillouin electron beam focusing system
equation \cite{22s,23s}, and $n=3$ gives the case of the Duffing oscillator %
\cite{6}.

An interesting particular case of the general anharmonic oscillator equation
Eq.~(\ref{1}) is the Ermakov-Pinney equation (EPE), which is a well-known
example of a nonlinear second order differential equation with important
physical applications (we refer the reader to \cite{6a,6aa} for an
historical development and an excellent review of the properties of the EPE
equation). The EPE is endowed with a wide range of physical applications,
including quantum cosmology \cite{7a}, dynamics of scalar field cosmologies
and the braneworld scenario \cite{8a}, quantum field theory \cite{9a,10a},
nonlinear elasticity \cite{11a}, nonlinear optics \cite{12a,13a},
description of the wavefunction of Bose-Einstein condensates (BEC) at the
mean-field level \cite{14a}, the envelope of the electric field in nonlinear
optics \cite{16a}, amongst others. In this context, the EPE provides an
effective description for the time-dependence of the relevant spatially
dependent field, typically being associated with its width both in the BEC %
\cite{17a,18a} and in optical settings \cite{19a}. The mathematical analysis
and structure of the EPE have been extensively discussed in \cite%
{20a,21a,22a,23a}.

Note that for generic values of the coefficients, Eq.~(\ref{1}) is
equivalent to a third order autonomous dynamical system, which generically
admits no closed form general solution \cite{7}. The mathematical properties
and applications of particular forms of Eq.~(\ref{1}) have been widely
investigated, such as, the partial integrability of the anharmonic
oscillator \cite{7}, the time-dependent driven anharmonic oscillator and its
adiabaticity properties \cite{8}, toroidal $p$-branes, anharmonic
oscillators and (hyper)elliptic solutions \cite{9}, conformal mappings and
other power series methods for solving ordinary differential equations \cite%
{10}, and the anharmonic oscillator in the context of the optimized basis
expansion \cite{11}. The Painlev\'{e} analysis of Eq.~(\ref{1}) was
performed in \cite{13}. Specific transformation properties of the anharmonic
oscillator were considered in \cite{12}, where an excellent review of the
Lie symmetries approach to Eq.~(\ref{1a}) can also be found.

The most general conditions on the functions $f_{1}$, $f_{2}$ and $f_{3}$,
for which Eq.~(\ref{1a}) may be integrable, as well as conditions for the
existence of Lie point symmetries, were obtained in \cite{12}.
Time-dependent first integrals were also constructed. The main results of %
\cite{12} are that if $n\notin \left\{ -3,-1,0,1\right\} $, then Eq.~(\ref%
{1a}) can be point transformed to an equation of the form $%
d^{2}X/dT^{2}+X^{n}\left( T\right) =0$, can be linearized as $%
d^{2}X/dT^{2}+k_{2}=0$, $k_{2}\in \Re \backslash {0}$, and it admits a
two-dimensional Lie point symmetry algebra.

It is the purpose of the present paper to obtain, by using the results of %
\cite{12}, some classes of exact solutions of the anharmonic oscillator Eq.~(%
\ref{1}) without the forcing term. The first solution is obtained by
considering a particular solution of the point transformed equation $%
d^{2}X/dT^{2}+X^{n}\left( T\right) =0$, equivalent to the initial anharmonic
oscillator equation. The integrability condition obtained in \cite{12} for
the anharmonic oscillator can be formulated in terms of a Riccati equation
for the $f_{1}(t)$ and $\frac{1}{f_{3}(t)}\frac{df_{3}}{dt}$ terms, respectively.
By imposing some specific constraints on the coefficients of the Riccati
equation, namely, by requiring that the Riccati equation can be reduced to a
Bernoulli equation, two distinct classes of exact solutions of the
anharmonic oscillator equation with zero forcing term are obtained. In the
analysis outlined below, we shall use the generalized Sundman
transformations $X\left( T\right) =F\left( t,x\right) $ and $dT=G\left(
t,x\right) dt$ \cite{14,15,16}. The latter have been widely applied in the
literature \cite{14,15}, namely, in the study of the mathematical properties
of the second order differential equations, and the third order differential
equation
\begin{equation}
\frac{d^{3}x}{dt^{3}}+h_{1}\left( t\right) \frac{d^{2}x}{dt^{2}}+h_{2}\left(
t\right) \frac{dx}{dt}+h_{3}\left( t\right) x+h_{4}\left( t\right) =0.
\end{equation}

The present paper is organized as follows. Three distinct classes of general
solutions of Eq.~(\ref{1}) without the forcing term, which explicitly depict the time
evolution of the anharmonic oscillator, are presented in Section \ref%
{sect2_1}. We discuss and conclude our results in Section \ref{sect3}.

\section{Exact integrability cases for the anharmonic oscillator}

\label{sect2_1}

In the present Section, by starting from the integrability condition of the
anharmonic oscillator equation obtained in \cite{12}, we obtain three cases
of exact integrability of the anharmonic oscillator without forcing.

\subsection{The integrability condition for the anharmonic oscillator}

\label{sect2}

In the following we assume that the forcing term $f_{4}( t) $ vanishes in
Eq.~(\ref{1}). Hence the latter takes the form
\begin{equation}
\frac{d^{2}x}{dt^{2}}+f_{1}\left( t\right) \frac{dx}{dt}+f_{2}\left(
t\right) x+f_3(t) x^{n}=0.  \label{1a}
\end{equation}

An integrability condition of Eq.~(\ref{1a}) can be formulated as the
following:

\textbf{Theorem} \cite{12}. If and only if $n\notin \left\{
-3,-1,0,1\right\} $, and the coefficients of Eq.~(\ref{1a}) satisfy the
differential condition
\begin{equation}
f_{2}\left( t\right) =\frac{1}{n+3}\frac{1}{f_{3}(t)}\frac{d^{2}f_{3}}{dt^{2}%
}-\frac{n+4}{\left( n+3\right) ^{2}}\left[ \frac{1}{f_{3}(t)}\frac{df_{3}}{dt%
}\right] ^{2}+\frac{n-1}{\left( n+3\right) ^{2}}\left[ \frac{1}{f_{3}(t)}%
\frac{df_{3}}{dt}\right] f_{1}\left( t\right) +\frac{2}{n+3}\frac{df_{1}}{dt}%
+\frac{2\left( n+1\right) }{\left( n+3\right) ^{2}}f_{1}^{2}\left( t\right) ,
\label{5c}
\end{equation}%
with the help of the pair of transformations
\begin{eqnarray}
X\left( T\right) &=&Cx\left( t\right) f_{3}^{\frac{1}{n+3}}\left( t\right)
e^{\frac{2}{n+3}\int^{t}f_{1}\left( \phi \right) d\phi },  \label{3b} \\
T\left( x,t\right) &=&C^{\frac{1-n}{2}}\int^{t}f_{3}^{\frac{2}{n+3}}\left(
\xi \right) e^{\left( \frac{1-n}{n+3}\right) \int^{\xi }f_{1}\left( \phi
\right) d\phi }d\xi ,  \label{4a}
\end{eqnarray}%
where $C$ is an arbitrary constant, Eq.~(\ref{1a}) can be point transformed
into the second order differential equation for $X\left( T\right) $,
\begin{equation}
\frac{d^{2}X}{dT^{2}}+X^{n}\left( T\right) =0.  \label{2a}
\end{equation}

The general solution of Eq.~(\ref{2a}) is given by
\begin{equation}
T=T_{0}+\epsilon \int \frac{dX}{\sqrt{2\left( C_{0}-\frac{X^{n+1}}{n+1}%
\right) }}, \qquad n\neq -1,  \label{T}
\end{equation}
where $T_{0}$ and $C_{0}$ are arbitrary constants of integration. For
convenience, we have denoted $T=T_{\pm }$, $C_{0}=C_{0\pm }$ $T_{0}=T_{0\pm
} $ and $\epsilon =\pm $.

By substituting the integrability condition given by Eq.~(\ref{5c}) into
Eq.~(\ref{1a}), we obtain the following integrable differential equation
\begin{eqnarray}
&&\frac{d^{2}x}{dt^{2}}+f_{1}\left( t\right) \frac{dx}{dt}+\Bigg\{ \frac{1}{%
n+3} \frac{1}{f_3(t)}\frac{d^{2}f_{3}}{dt^{2}} -\frac{n+4}{\left( n+3\right)
^{2}}\left[ \frac{1}{f_3(t)}\frac{df_{3}}{dt}\right] ^{2} +  \notag \\
&&\frac{n-1}{\left( n+3\right) ^{2}}\left[ \frac{1}{f_3(t)}\frac{df_{3}}{dt}%
\right] f_{1}\left( t\right) +\frac{2}{n+3}\frac{df_{1}}{dt}+\frac{2\left(
n+1\right) }{\left( n+3\right) ^{2}}f_{1}^{2}\left( t\right) \Bigg\} %
x+f_{3}\left( t\right) x^{n}=0, n\notin \left\{-3,-1,0,1\right\}.
\label{15m}
\end{eqnarray}

\subsection{A particular exact solution for the anharmonic oscillator
equation}

The general solution of Eq.~(\ref{2a}) can be given as
\begin{equation}
T=T_{0}+\frac{\epsilon }{C_{0}}X\sqrt{\frac{C_{0}\left( n+1\right) -X^{n+1}}{%
2\left( n+1\right) }}\,_{2}F_{1}\left[ 1,\frac{n+3}{2\left( n+1\right) };%
\frac{n+2}{n+1};\frac{X^{n+1}}{C_{0}(n+1)}\right] , \qquad   n\neq -1,
\end{equation}%
where $_{2}F_{1}(a,b;c;d)$ is the hypergeometric function. A particular
solution of Eq.~(\ref{2a}) is given by
\begin{equation}
X\left( T\right) =\left[ \epsilon \left( T-T_{0}\right) \right] ^{\frac{2}{%
1-n}}\left[ -\frac{\left( n-1\right) ^{2}}{2\left( n+1\right) }\right] ^{%
\frac{1}{1-n}},  \label{X2}
\end{equation}%
where we have defined $X\left( T\right) =X_{\pm }\left( T\right) $, and we have taken
the arbitrary integration constant as zero, $C_{0}=0$. In order to have a real value of the displacement $x(t)$ one must impose the condition  $n<-1$ on the anharmonicity exponent $n$. From
Eqs.~(\ref{3b}) and (\ref{X2}), we obtain the result
\begin{equation}
x\left( t\right) =\frac{1}{C}\left[ \epsilon \left( T-T_{0}\right) \right] ^{%
\frac{2}{1-n}}\left[ -\frac{\left( n-1\right) ^{2}}{2\left( n+1\right) }%
\right] ^{\frac{1}{1-n}}f_{3}^{-\frac{1}{n+3}}\left( t\right) e^{-\frac{2}{%
n+3}\int^{t}f_{1}\left( \phi \right) d\phi },  \label{X1}
\end{equation}%
where we have denoted $x\left( t\right) =x_{\pm }\left( t\right) $, for
simplicity.

By inserting Eq. (\ref{4a}) into Eq. (\ref{X1}) yields the general solution
of Eq. (\ref{15m}) describing the time evolution of anharmonic oscillator.
Therefore we have obtained the following:

\textbf{Corollary 1}. The anharmonic oscillator equation Eq.~(\ref{15m}) has
the particular solution
\begin{equation}
x\left( t\right) =x_{0}\left[ C^{\frac{1-n}{2}}\int^{t}f_{3}^{\frac{2}{n+3}%
}\left( \xi \right) e^{\left( \frac{1-n}{n+3}\right) \int^{\xi }f_{1}\left(
\phi \right) d\phi }d\xi -T_{0}\right] ^{\frac{2}{1-n}}f_{3}^{-\frac{1}{n+3}%
}\left( t\right) e^{-\frac{2}{n+3}\int^{t}f_{1}\left( \phi \right) d\phi
},n<-1,  \label{X2k}
\end{equation}
where we have defined $x_{0}=C^{-1}\left[ -\frac{\left( n-1\right) ^{2}}{%
2\left( n+1\right) }\right] ^{\frac{1}{1-n}}$.

\subsection{Second integrability case for the anharmonic oscillator equation}

Now, by rearranging the terms of Eq.~(\ref{5c}) yields the following Riccati
equation for $f_{1}\left( t\right) $ given by
\begin{equation}
\frac{df_{1}}{dt}=a\left( t\right) +b\left( t\right) f_{1}\left( t\right)
+c\left( t\right) f_{1}^{2}\left( t\right) ,  \label{6a}
\end{equation}%
where the coefficients $a(t)$, $b(t)$ and $c(t)$ are defined as
\begin{eqnarray}
a\left( t\right) &=&\frac{3+n}{2}f_{2}\left( t\right) -\frac{1}{2f_{3}\left(
t\right) }\frac{d^{2}f_{3}}{dt^{2}}+\frac{4+n}{2\left( 3+n\right) }\left[
\frac{1}{f_{3}\left( t\right) }\frac{df_{3}}{dt}\right] ^{2},  \label{7a} \\
b\left( t\right) &=&\frac{1-n}{2\left( 3+n\right) }\left[ \frac{1}{%
f_{3}\left( t\right) }\frac{df_{3}}{dt}\right] ,  \label{8a} \\
c\left( t\right) &=&-\frac{1+n}{3+n}.  \label{9a}
\end{eqnarray}

We consider now that the coefficient $a\left( t\right) $ of Eq.~(\ref{6a})
vanishes, so that Eq. (\ref{7a}) can be written as
\begin{equation}
f_{2}\left( t\right) =\frac{1}{3+n}\frac{1}{f_{3}\left( t\right)} \frac{%
d^{2}f_{3}}{dt^{2}} -\frac{4+n}{\left( 3+n\right) ^{2}}\left[ \frac{1}{%
f_{3}\left( t\right)} \frac{df_{3}}{dt}\right] ^{2}.  \label{10a}
\end{equation}
Hence the Riccati Eq. (\ref{6a}) becomes a Bernoulli type equation, with the
general solution given by
\begin{equation}
f_{1}\left( t\right) =\frac{f_{3}^{\frac{1-n}{2\left( 3+n\right) }}\left(
t\right) }{C_{1}+\frac{1+n}{3+n}\int^{t}f_{3}^{\frac{1-n}{2\left( 3+n\right)
}}\left( \phi \right) d\phi },  \label{11a}
\end{equation}
where $C_{1}$ is an arbitrary constant of integration.

By substituting Eqs. (\ref{10a}) and (\ref{11a}) into Eq. (\ref{1a}), the
latter yields the following differential equation
\begin{equation}
\frac{d^{2}x}{dt^{2}}+\left[ \frac{f_{3}^{\frac{1-n}{2\left( 3+n\right) }%
}\left( t\right) }{C_{1} +\frac{1+n}{3+n}\int^{t}f_{3}^{\frac{1-n}{2\left(
3+n\right) }}\left( \phi \right) d\phi }\right] \frac{dx}{dt} +\left\{ \frac{%
1}{3+n}\frac{1}{f_{3}\left( t\right)} \frac{d^{2}f_{3}}{dt^{2}} -\frac{4+n}{%
\left( 3+n\right) ^{2}}\left[ \frac{1}{f_{3}\left( t\right)} \frac{df_{3}}{dt%
}\right] ^{2} \right\} x+f_{3}\left( t\right) x^{n}=0.  \label{12a}
\end{equation}

Therefore we have obtained the following:

\textbf{Corollary 2}. The general solution of Eq.~(\ref{12a}), describing
the time evolution of the anharmonic oscillator, is given by
\begin{eqnarray}
x\left( t\right) &=&x_{0}\left[ C^{\frac{1-n}{2}}\int^{t}f_{3}^{\frac{2}{n+3}%
}\left( \xi \right) e^{\left( \frac{1-n}{n+3}\right) \int^{\xi }\left[ \frac{%
f_{3}^{\frac{1-n}{2\left( 3+n\right) }}\left( \phi \right) }{C_{1}+\frac{1+n%
}{3+n}\int^{\phi }f_{3}^{\frac{1-n}{2\left( 3+n\right) }}\left( \psi \right)
d\psi }\right] d\phi }d\xi -T_{0}\right] ^{\frac{2}{1-n}}\times  \notag \\
&&\times f_{3}^{-\frac{1}{n+3}}\left( t\right) e^{-\frac{2}{n+3}\int^{t}%
\left[ \frac{f_{3}^{\frac{1-n}{2\left( 3+n\right) }}\left( \phi \right) }{%
C_{1}+\frac{1+n}{3+n}\int^{\phi }f_{3}^{\frac{1-n}{2\left( 3+n\right) }%
}\left( \xi \right) d\xi }\right] d\phi },  \qquad n\notin \left\{-3,-1,0,1\right\}.
\label{13a}
\end{eqnarray}

\subsection{Third integrability case for the anharmonic oscillator equation}

Now, by introducing a new function $u\left( t\right) $ defined as
\begin{equation}
u\left( t\right) =\frac{1}{f_{3}\left( t\right) }\frac{df_{3}}{dt},
\end{equation}
or, equivalently,
\begin{equation}
f_{3}\left( t\right) =f_{03}e^{\int^{t}u\left( \phi \right) d\phi },
\end{equation}
where $f_{03}$ is an arbitrary constant of integration, after substituting $%
u\left( t\right) $ into Eq.~(\ref{5c}) yields the following Riccati equation
for $u\left( t\right) $, given by
\begin{equation}
\frac{du}{dt}=a_{1}\left( t\right) +b_{1}\left( t\right) u\left( t\right)
+c_{1}\left( t\right) u^{2}\left( t\right) ,  \label{6k}
\end{equation}%
where the coefficients are defined as
\begin{eqnarray}
a_{1}\left( t\right) &=&\left( 3+n\right) f_{2}\left( t\right) -\frac{%
2\left( 1+n\right) }{3+n}f_{1}^{2}\left( t\right) -2\frac{df_{1}}{dt},
\label{7k} \\
b_{1}\left( t\right) &=&\frac{1-n}{3+n}f_{1}\left( t\right) ,  \label{8k} \\
c_{1}\left( t\right) &=&\frac{1}{3+n}.  \label{9k}
\end{eqnarray}

We consider that the coefficient $a_1\left( t\right) $ of Eq.~(\ref{6k})
vanishes, as before, so that Eq.~(\ref{7k}) can be written as
\begin{equation}
f_{2}\left( t\right) =\frac{2\left( 1+n\right) }{\left( 3+n\right) ^{2}}%
f_{1}^{2}\left( t\right) +\frac{2}{\left( 3+n\right) }\frac{df_{1}}{dt}.
\label{10m}
\end{equation}

Then the Riccati Eq. (\ref{6k}) becomes a Bernoulli type equation, with the
general solution given by
\begin{equation}
u\left( t\right) =\frac{e^{\frac{1-n}{3+n}\int^{t}f_{1}\left( \phi \right)
d\phi }}{C_{2}-\frac{1}{3+n}\int^{t}e^{\frac{1-n}{3+n}\int^{\xi }f_{1}\left(
\phi \right) d\phi }d\xi },  \label{11m}
\end{equation}
where $C_{2}$ is an arbitrary constant of integration. Thus, the coefficient
$f_{3}\left( t\right) $ of Eq. (\ref{1a}) is readily given by
\begin{equation}
f_{3}\left( t\right) =f_{03}e^{\int^{t}\left[ \frac{e^{\frac{1-n}{3+n}%
\int^{\psi }f_{1}\left( \phi \right) d\phi }}{C_{2}-\frac{1}{3+n}\int^{\psi
}e^{\frac{1-n}{3+n}\int^{\xi }f_{1}\left( \phi \right) d\phi }d\xi }\right]
d\psi }.  \label{12k}
\end{equation}

By substituting Eqs.~(\ref{10m}) and (\ref{12k}) into Eq. (\ref{1a}), the
latter yields the following differential equation
\begin{equation}
\frac{d^{2}x}{dt^{2}}+f_{1}\left( t\right) \frac{dx}{dt}+\left[ \frac{%
2\left( 1+n\right) }{\left( 3+n\right) ^{2}}f_{1}^{2}\left( t\right) +\frac{2%
}{\left( 3+n\right) }\frac{df_{1}}{dt}\right] x+f_{03}e^{\int^{t}\left[
\frac{e^{\frac{1-n}{3+n}\int^{\psi }f_{1}\left( \phi \right) d\phi }}{C_{2}-%
\frac{1}{3+n}\int^{\psi }e^{\frac{1-n}{3+n}\int^{\xi }f_{1}\left( \phi
\right) d\phi }d\xi }\right] d\psi }x^{n}=0.  \label{13k}
\end{equation}

Therefore we have obtained the following:

\textbf{Corollary 3}. The general solution of Eq. (\ref{13k}), describing
the time evolution of an anharmonic oscillator, is given by
\begin{eqnarray}
x\left( t\right) &=&x_{0}f_{03}^{-\frac{1}{3+n}}\left[ C^{\frac{1-n}{2}%
}f_{03}^{\frac{2}{3+n}}\int^{t}e^{\frac{2}{3+n}\int^{\xi }\left[ \frac{e^{%
\frac{1-n}{3+n}\int^{\psi }f_{1}\left( \phi \right) d\phi }}{C_{2}-\frac{1}{%
3+n}\int^{\psi }e^{\frac{1-n}{3+n}\int^{\rho }f_{1}\left( \phi \right) d\phi
}d\rho }+\frac{1-n}{2}f_{1}\left( \psi \right) \right] d\psi }d\xi -T_{0}%
\right] ^{\frac{2}{1-n}}\times  \notag \\
&& \times \; e^{-\frac{1}{3+n}\int^{t}\left[ \frac{e^{\frac{1-n}{3+n}%
\int^{\psi }f_{1}\left( \phi \right) d\phi }}{C_{2}-\frac{1}{3+n}\int^{\psi
}e^{\frac{1-n}{3+n}\int^{\rho }f_{1}\left( \phi \right) d\phi }d\rho }%
+2f_{1}\left( \psi \right) \right] d\psi },  \qquad n\notin \left\{-3,-1,0,1\right\}.
\label{15k}
\end{eqnarray}

\section{Conclusions}

\label{sect3}

In the limit of a small function $X\left( T\right) $, and by assuming that
the constant $n$ is large, $n\rightarrow +\infty $, in view of Eq.~(\ref{2a}%
) we obtain a linear relation between $X\left( T\right) $ and $T\left(
t\right) $, given by
\begin{equation}
X\left( T\right) =\epsilon \sqrt{2C_{0}}\left( T-T_{0}\right) .  \label{16k}
\end{equation}%
With the help of Eqs.~(\ref{3b}) and (\ref{4a}), the approximate solution of
Eq.~(\ref{1a}), describing the time evolution of anharmonic oscillator is
given by
\begin{equation}
x\left( t\right) \approx \frac{\epsilon \sqrt{2C_{0}}}{C}\left[ C^{\frac{1-n%
}{2}}\int^{t}f_{3}^{\frac{2}{n+3}}\left( \xi \right) e^{\left( \frac{1-n}{n+3%
}\right) \int^{\xi }f_{1}\left( \phi \right) d\phi }d\xi -T_{0}\right]
f_{3}^{-\frac{1}{n+3}}\left( t\right) e^{-\frac{2}{n+3}\int^{t}f_{1}\left(
\phi \right) d\phi }.  \label{17k}
\end{equation}%
With this approximate solution, once the functions $f_{1}(t)$ and $f_{3}(t)$
are given, one can study the time evolution of the anharmonic oscillator for
small $X\left( T\right) $ and for a very large anharmonicity exponent $n$.

in the present paper, by extending the results of \cite{12}, where the first integral of Eq.~(\ref%
{1a}) was obtained, we have obtained three classes of
exact general solutions of Eq.~(\ref{1a}), by explicitly showing that the
Theorem obtained in \cite{12} is very useful for obtaining the explicit
general solutions of the anharmonic oscillator type second order
differential equations.

In order to have real solutions, the general solutions Eqs. (\ref{X2k}), (%
\ref{13a}) and (\ref{15k}) of the second order differential Eqs. (\ref{15m}%
), (\ref{12a}) and (\ref{13k}), respectively, must obey the condition $n<-1$%
, thus leading to an anharmonic term of the form $f_3(t)/x^n$, $n>0$. Such a
term may be singular at $x=0$. Note that in \cite{20s}, the author has used
the degree theory to remove a technical assumption in the non-resonance
results in \cite{21s} and to obtain a complete set of the non-resonance
conditions for differential equations with repulsive singularities. In doing
so, a nice relation between the Hill's equation $\frac{d^{2}x}{dt^{2}}+\eta
\left( t\right) x\left( t\right) =0$ and the EPE was established. This
relation itself is useful in studying the stability of periodic solutions of
Lagrangian systems of degree of freedom of $3/2$.

It is well-known that the second order ordinary differential equations with
anharmonic term of the form $f_{3}(t)/x^{n}$ entail many problems in the
applied sciences. Some examples are the Brillouin focusing system, and the
motion of an atom near a charged wire. The Brillouin focusing system can be
described by the second order differential equation
\begin{equation}
\frac{d^{2}x}{dt^{2}}+\alpha \left( 1+\cos t\right) x\left( t\right) =\frac{%
\beta }{x\left( t\right) },
\end{equation}%
where $\alpha $ and $\beta $ are positive constants. In the context of
electronics, this differential equation governs the motion of a magnetically
focused axially symmetric electron beam under the influence of a Brillouin
flow, as shown in \cite{22s}. From the mathematical point of view, this
differential equation is a singular perturbation of a Mathieu equation
\begin{equation}
\frac{d^{2}x}{dt^{2}}+\left( a-2q\cos 2t\right) x\left( t\right) =0,
\end{equation}%
where $a$ and $q$ are arbitrary constants. Existence and uniqueness of
elliptic periodic solutions of the Brillouin electron beam focusing system
has been discussed in \cite{23s}. Hence, the results obtained in the present
paper could open the possibility of obtaining some exact solutions of
non-linear differential equations of scientific or technological interest.

\end{document}